\documentclass[10pt,a4paper,oneside,onecolumn]{article}
\usepackage[]{fontenc}
\usepackage[dvips]{epsfig}

\makeatletter
\def\LyX{L\kern-.1667em\lower.25em\hbox{Y}\kern-.125emX\spacefactor1000}%
\newcommand{\lyxtitle}[1] {\thispagestyle{empty}
\global\@topnum\z@
\section*{\LARGE \centering \sffamily \bfseries \protect#1 }
}

\newcommand{\lyxletterstyle}{
\setlength\parskip{0.7em}
\setlength\parindent{0pt}
}
\newcommand{\lyxaddress}[1]{
\par {\raggedright #1 
\vspace{1.4em}
\noindent\par}
}

\makeatother

\lyxletterstyle
\begin{document}

{\bfseries \large \hfill{}Barrier Perturbation Induced ``Superarrivals''
 and \hfill{} \par}

{\bfseries \large \hfill{}``Nonlocality'' in a Time-Evolving Wave
 Packet\hfill{} \par}

\( \smallskip  \)

\large \hfill{}\normalsize Somshubhro Bandyopadhyay and Dipankar
Home\footnote{
email: dhom@boseinst.ernet.in
} \hfill{}

\lyxaddress{\hfill{}\small Department of Physics, Bose Institute 93/1 A. P.
C Road Calcutta -700009, INDIA\normalsize \hfill{} }

\( \smallskip  \)

{\small We compute the \em time evolving \em probability of a Gaussian
wave packet to be reflected from a rectangular potential barrier which
is perturbed by \em reducing \em its height. A time interval is found
during which this probability of reflection is \em enhanced \em (``superarrivals'')
compared to the unperturbed case. Such a time evolving reflection
probability implies that the effect of perturbation propagates across
the wave packet \em faster \em than its group velocity - a curious
form of ``nonlocality.'' \par}

\( \smallskip  \)

In recent years a number of interesting investigations have been reported
on wave- packet dynamics in quantum-well systems [1-6]. In this paper
we study the wave packet dynamics from a new perspective. The reflection/transmission
probabilities for the scattering of wave packets by various obstacles
are usually considered from static or unperturbed potential barriers.
Generally the \em time-independent \em (asymptotic) values attained
after a complete time evolution are calculated. Here we point out
the striking effects that occur \em during \em the time evolution
by considering dynamics of wave packet scattering from a \em \em barrier
whose height is reduced to zero \em well before \em the asymptotic
value of reflection probability is reached.

For an unperturbed barrier the reflection probability for an initially
localized wave packet \( \psi \left( x,t=0\right)  \) is calculated by considering the wave packet
as a superposition of plane waves and by writing

\begin{equation}
\label{1}
\left| R_{0}\right| ^{2}=\int \left| \phi \left( p\right) \right| ^{2}\left| R\left( p\right) \right| ^{2}dp
\end{equation}

where \( \left| R\left( p\right) \right| ^{2} \) is the reflection probability corresponding to the plane wave
component \( exp(ipx) \) and \( \phi (p) \) is Fourier transform of the initial wave packet
\( \psi (x,t=0) \). Since a wave packet evolves in time, \( \left| R_{0}\right| ^{2} \) defined by Eq. (1) denotes
the \em time-independent \em value of reflection probability pertaining
to a wave packet, this value being attained in the asymptotic limit
(\( t_{\infty } \)) of the time evolution. Thus \( \left| R_{0}\right| ^{2} \) can be expressed in the following
form

\begin{equation}
\label{2}
\left| R_{0}\right| ^{2}=\int ^{x\prime }_{-\infty }\left| \psi \left( x,t_{\infty }\right) \right| ^{2}dx
\end{equation}

where \( \psi \left( x,t_{\infty }\right)  \) is asymtotic form of the wave packet attained by evolving
from \( \psi (x,t=0) \) and by being scattered from a rectangular potential barrier
of finite height and width. Note that \( x\prime  \) lies at a left edge of the
initial profile of the wave packet such that \( \int ^{x\prime }_{-\infty }\left| \psi \left( x,t=0\right) \right| ^{2}dx \) is negligible (see
Figure 1). At any instant \em before \em the constant value \( \left| R_{0}\right| ^{2} \) is attained,
the time evolving reflection probability in the region \( -\infty <x\leq x\prime  \) is given
by

\begin{equation}
\label{3}
\left| R(t)\right| ^{2}=\int ^{x\prime }_{-\infty }\left| \psi \left( x,t\right) \right| ^{2}dx
\end{equation}

Now suppose that during the time evolution of this wave packet the
barrier is perturbed by \em reducing \em its height to zero within
a very \em short \em but \em finite \em interval of time. Here by
``short'' time interval we mean that it is small compared to the time
taken by the reflection probability to attain its asymptotic value
\( \left| R_{0}\right| ^{2} \). We compute effects \em \em of this ``sudden'' perturbation on \( \left| R(t)\right| ^{2} \).
The salient features of our findings are as follows:

(a) A finite time interval is found during which \( \left| R(t)\right| ^{2} \) shows a surprising
\em enhancement \em (we call this effect ``superarrivals'') in the perturbed
case even though the barrier height is reduced. This time interval
and the amount of enhancement depend on the time over which the barrier
height is made zero. \\
(b) The way the computed reflection probability evolves implies an
\em incompatibility \em with a form of ``locality condition'' which
is inferred from a commonly used ``particle picture'' which associates
the mean ``velocity'' of a ``particle'' with group velocity of the corresponding
wave packet. In particular, such a ``picture'' is \em phenomenologically
\em useful in the context of design and interpretation of ``single
particle'' interference experiments using neutrons/electrons [7,8].
The type of quantum nonlocality thus exhibited involves an \em action
at a distance \em entailing a \em global \em effect on the wave packet
induced by a local perturbation. The resulting action propagates across
the wave packet at a finite speed which is \em greater \em than the
group velocity - a distinctly nonclassical behaviour.

In order to demonstrate the above features let us begin by writing
the initial wave packet (in units of \( \hbar =1 \) and \( m=1/2 \), this choice of units
being convenient for numerical computation) in the form

\begin{equation}
\label{4}
\psi \left( x,t=0\right) =\frac{1}{\left[ 2\pi \left( \sigma _{0}\right) ^{2}\right] ^{1/4}}exp\left[ -\frac{\left( x-x_{0}\right) ^{2}}{4\sigma ^{2}_{0}}+ip_{0}x\right] 
\end{equation}

which describes a packet of width \( \sigma _{0} \) centered around \( x=x_{0} \) with its peak
moving with a group velocity \( 2p_{0}=\frac{\left\langle p\right\rangle }{m} \) towards a rectangular potential barrier.
The point \( x_{0} \) is chosen such that \( \psi \left( x,t=0\right)  \) has a negligible overlap with the
barrier. The expectation value of energy \( (E) \) of the wave packet is given
by \( p^{2}_{0}+\frac{1}{4}\sigma ^{-2}_{0} \).\\
For the purpose of computing \( \left| R(t)\right| ^{2} \) given by Eq. (3) the time dependent
Schrodinger equation is solved by using the numerical methods as developed
by Goldberg , Schey and Schwartz [9]. In this treatment the parameters
are chosen in a way ensuring the spreading of packet to be negligible
during the scattering process so that it doesn't mask the effects
of interest. Here we choose \( x_{0}=1.2 \), \( \sigma _{0}=0.05/\sqrt{2} \) and \( p_{0}=50\pi  \). The barrier is centered around
\( x_{c}=1.5 \) with width = 0.064. Height of the barrier (V) before perturbation
is chosen to be \( V=2E \) . This choice satisfies the following criteria :

(1) \( V \) is such that the reflection probability is very close to \( 1 \) since
we are interested only in the reflection probability.

(2) At the same time \( V \) is chosen not to be too large . This is in
order to ensure that reduction of the barrier height need not be too
fast.

\( \left| R(t)\right| ^{2} \) is computed according to Eq. (3) by taking \( x^{\prime }=x_{0}-3\sigma _{0}/\sqrt{2} \) . The computed evolution
of \( \left| R(t)\right| ^{2} \) corresponds to the building up of reflected particles with time.
More precisely, it means that a detector located within the region
\( -\infty <x<x^{\prime } \) measures \( \left| R(t)\right| ^{2} \) by registering the reflected particles arriving in that
region \em up to \em various instants. \\
First, we compute \( \left| R(t)\right| ^{2} \) for the wave packet scattered from a \em static
\em barrier \( V=2E \). The relevant curve is shown in Figure 2 which tends
towards a time-independent value which is the stationary state reflection
probability \( \left| R_{0}\right| ^{2} \) given by Eq (2); this is numerically verified to be
equivalent to the expression for \( \left| R_{0}\right| ^{2} \) given by Eq. (1). We then proceed
to study the consequences of reducing the barrier height from \( V=2E \) to
V = 0. The time evolution of \( \left| R(t)\right| ^{2} \) in the perturbed cases is found to
show a number of intreresting features . \\
In all the cases we study, the potential \( V \) goes to zero linearly within
a \em switching off time \em \( \epsilon  \) around \( t=t_{p} \) chosen to be \( 8\times 10^{-4} \) (note that
numbers denoting the various instants are in terms of time steps;
for example, \( t=8\times 10^{-4} \) corresponds to 400 time steps). Here \( \epsilon \ll t_{0} \), \( t_{0} \) being the
time required for \( \left| R(t)\right| ^{2} \) to attain the asymptotic value \( \left| R_{0}\right| ^{2} \). This short time
span \( \epsilon  \) over which the perturbation takes place is thus given by \( \left[ t_{p}-\frac{\epsilon }{2},t_{p}+\frac{\epsilon }{2}\right]  \).
Profile of the wave packet at \( t_{p}=8\times 10^{-4} \) is shown in Figure 3. Note that at
\em that \em instant the overlap of the wave packet with the barrier
is significant. \\
Figure 4 shows the evolution of \( \left| R(t)\right| ^{2} \)for various values of \( \epsilon  \). Different
\( \epsilon  \) correspond to different \( N \) where \( N \) is the number of time steps involved
in computing the reduction of \( V \) from \( 2E \) to \em \em zero, \( \frac{\epsilon }{N} \) being the
magnitude of each time step. Varying \( \epsilon  \) signifies changing the time
span over which the barrier height goes to zero which in turn means
different rates of reduction. We now compare the computed \( \left| R(t)\right| ^{2} \) for the
particular case N=2 (denoted by \( \left| R_{p}(t)\right| ^{2} \)) with that calculated for a static
barrier (denoted by \( \left| R_{s}(t)\right| ^{2} \)). This comparison is shown in Figure 5 which
reveals that \\

\begin{equation}
\label{5}
\left| R_{p}(t)\right| ^{2}=\left| R_{s}(t)\right| ^{2}\quad \quad \quad t\leq t_{d}
\end{equation}

\begin{equation}
\label{6}
\left| R_{p}(t)\right| ^{2}>\left| R_{s}(t)\right| ^{2}\quad \quad \quad t_{d}<t\leq t_{c}
\end{equation}
 
\begin{equation}
\label{7}
\left| R_{p}(t)\right| ^{2}<\left| R_{s}(t)\right| ^{2}\quad \quad \quad t>t_{c}
\end{equation}

where \( t_{p} \) is the instant around which the perturbation takes place,
\( t_{c} \) is the instant when the two curves cross each other, and \( t_{d} \) is the
time  from which the curve corresponding to the perturbed case starts
\em deviating \em from that in the unperturbed case. Here \( t_{c}>t_{d}>t_{p} \). \\
Let us now focus on a striking feature embodied in the inequality
(6). As the barrier height is made zero, one does not expect at any
time an increase in the reflected particle flux compared to that in
the unperturbed case. Nevertheless, the inequality (6) shows that
there is a finite time interval \( \Delta t=\left[ t_{d},t_{c}\right]  \) during which the probability of
finding a reflected particle is \em more \em (``superarrivals'') in
the perturbed case than when the barrier is left unperturbed (see
Figure 5). A detector placed in the region \( x<x' \) would therefore register
\em more \em counts during this time interval \( \Delta t \) \em even though \em the
barrier height had been reduced to zero prior to that. It has been
checked that this effect of ``superarrivals'' occur for other values
of \( N \) (or, \( \epsilon  \)) as well; see Figure 4. 

Figure 4 also reveals that the probability of ``superarrivals'' (i.e.,
\em the enhancement \em of reflection probability) depends on \( N \) (or,
\( \epsilon  \)). The maximum enhancement takes place for \( N=2 \) and the amount of enhancement
decreases with increasing \( N \). But interestingly there is no appreciable
change in the magnitude of the time interval \( \Delta t \) over which this enhancement
occurs. In order to have a \em quantitative \em measure of ``superarrivals''
we define the parameter \( \eta  \) given by

\begin{equation}
\label{8}
\eta =\frac{I_{p}-I_{s}}{I_{s}}
\end{equation}

where the quantities \( I_{p} \) and \( I_{s} \) are defined with respect to \( \Delta t \) during
which ``superarrivals'' occur

\begin{equation}
\label{9a}
I_{p}=\int _{\Delta t}\left| R_{p}(t)\right| ^{2}dt
\end{equation}

\begin{equation}
\label{9b}
I_{s}=\int _{\Delta t}\left| R_{s}(t)\right| ^{2}dt
\end{equation}

The relevant numerical results are displayed in  Table 1 and variation
of \( \eta  \) with \( N \) (or, \( \epsilon  \)) for different cases of perturbation are shown
in Figure 6. These results are summarised below:

(a) There exists a \em finite \em time interval \( \Delta t \)
 during which an
\em increase \em in the reflection probability (``superarrivals'') occurs
for the perturbed cases relative to the unperturbed situation. 

(b) The time interval \( \Delta t \) over which this probability enhancement takes
place is \em not \em sensitive to the span of the \em switching off
time \em \( \epsilon  \) within which the barrier height is reduced to zero.

(c) Magnitude of this probability enhancement falls off \em linearly
\em with increasing \( \epsilon  \). 

We also note that both \( \Delta t \) and ``superarrivals'' given by \( \eta  \) depend on
the \em instant \( t_{p} \) \em around which the barrier is switched off. Hence
\em choice \em of this instant for demonstrating the above effects
needs to be appropiate. From the profiles of wave packet corresponding
to different times of perturbation \( t_{p} \) (as shown in Figure 6) it is
seen that ``superarrivals'' is appreciable in cases where the wave packet
has some significant overlap with the barrier \em during \em its switching
off. What \em optimal \em condition determining the choice of  \( t_{p} \) maximises
``superarrivals'' needs to be clarified. In particular, there could
be choice(s) of \( t_{p} \) for which ``superarrivals'' is more than what is obtained
in the present work.Further, it should be interesting to investigate
whether any \em lower bound \em of \( t_{p} \) exists prior to which switching
off the barrier does not give rise to ``superarrivals''.

\em Quantum Nonlocality.- \em We now proceed to discuss in what sense
the computed time development of \( \left| R(t)\right| ^{2} \) in the perturbed cases mentioned
above entail an \em incompatibility \em with a certain form of locality
condition. This locality condition is formulated in terms of a ``particle
picture'' which is commonly used in interpreting wave packet behaviour.
In particular, such a ``picture'' is phenomenologically motivated, being
used in analyzing the results of neutron/electron interferometric
experiments [7,8] performed in the region of so-called ``self interference''
where only one ``particle'' is required to be present inside the device
at a time. In order to ensure this condition it is necessary to associate
a mean ``velocity'' with an individual neutron/electron so that the
time it stays inside the interferometer can be calculated and hence
suitably adjusted by varying the relevant parameters (such as the
rate of emission from a source). 

A crucial point is that this mean ``velocity'' is assumed to be the
group velocity (\( v_{g} \)) of the wave packet associated with the particle
[10]. The average time of transit \( \Delta T \) of a neutron/electron inside the
device is then estimated by using the relation \( \Delta T=L/v_{g} \) where \( L \) is the distance
travelled within the device. Such a relation characterizes the type
of ``particle picture'' which constitutes a crucial ingredient in designing
and interpreting the neutron/electron self interference experiments.
This ``particle picture'', if applied in the specific context of the
example discussed in the present paper, leads to the following Propositions.

\em Proposition 1: \em If a particle is detected at time \( t, \) it is inferred
to be reflected from the barrier at an \em earlier instant \em \( t-\frac{D}{v_{g}} \) where
\( D \) is the separation between the detector and the barrier. 

\em Proposition 2: \em Locally perturbing a barrier has \em no \em effect
on those particles \em already \em reflected from the barrier. 

Note that Proposition 2 is some form of \em locality condition \em which
assumes that detection probability of the particles reflected from
the barrier prior to its perturbation does \em not \em bear any signature
of the perturbation. On the basis of the above Propositions we now
proceed to derive the following constraint condition. 

In our computed cases the relevant perturbation (i.e., reduction of
the barrier height) commences from the instant \( t_{p}-\frac{\epsilon }{2} \). From Proposition
1 it follows that particles reflected from the barrier \em until \em this
particular instant are registered at a detector (placed at a distance
\( D \)) \em up to \em an instant \( \tau  \) which is given by \( \tau =(t_{p}-\frac{\epsilon }{2})+\frac{D}{v_{g}} \). The locality condition
(Proposition 2) therefore requires that the measured particle statistics
at this detector should \em not \em be affected by perturbation of
the barrier until the instant \( \tau  \). This means that the time evolving
reflection probability \( \left| R_{p}(t)\right| ^{2} \) in the perturbed cases is permitted by this
locality condition to deviate from the reflection probability \( \left| R_{s}(t)\right| ^{2} \) in
the static case \em only after \em the instant \( \tau  \). In other words,
if \( \left| R_{p}(t)\right| ^{2} \) and \( \left| R_{s}(t)\right| ^{2} \) are found to differ from the instant \( t_{d} \) , the locality condition
in the form of Proposition 2 requires
\begin{equation}
\label{10}
t_{d}>\tau 
\end{equation}

However, the results of our computations of \( \left| R_{p}(t)\right| ^{2} \) and \( \left| R_{s}(t)\right| ^{2} \) indicate a rather
strong violation of the inequality (10). Choosing \( D=0.375 \),\( v=2p_{0}=100\pi  \) and \( t_{p}=8\times 10^{-4} \) , the
results of our calculation are displayed in Table 2.

Comparing the values of \( \tau  \) with those of \( t_{d} \) as shown in Table 2 we find
that \( t_{d}<\tau  \) for all values of \( \epsilon  \). A clear violation of the locality condition
(10) is thus demonstrated. This form of quantum nonlocality is quite
distinct from the usual nonlocality [10] which is inferred from many
particle entangled states. We now elaborate a bit on the significance
of this new form of quantum nonlocality. 

What our computed results show is that a local change in potential
(in our specific case, a reduction of the barrier height) affects
a wave packet \em globally\em , the global effect being manifested
through time evolution of the packet. The action due to local perturbation
propagates across the wave packet at a finite speed say, \( v_{e} \) affecting
the time evolving reflection probability which can be measured at
different points. Thus a distant observer who records the growth of
reflection probability becomes aware of perturbation of the barrier
(occuring around an instant \( t_{p} \)) from the instant \( t_{d} \) when the time varying
reflection probability starts \em deviating \em from that measured
in the unperturbed case. Then \( v_{e} \) is given by 
\begin{equation}
\label{11}
v_{e}=\frac{D}{t_{d}-(t_{p}-\frac{\epsilon }{2})}
\end{equation}

From Eq. (12) it follows that the violation of locality condition
(10) implies 
\begin{equation}
\label{12}
v_{e}>v_{g}
\end{equation}

i.e., the effect caused by reducing the barrier height travels across
the wave packet at a speed \em exceeding \em the packet's group velocity.
This is an intrinsically \em nonclassical \em ``action at a distance''
which is manifested even when spreading of the wave packet is ensured
to be negligibly \em small\em . \\
In general, \( v_{e} \) depends on \( D,t_{p,}t_{d,} \) and \( \epsilon  \). However, \( t_{d} \) is essentially determined
by \( \epsilon  \) for a given \( t_{p} \). Hence for fixed values of \( D \) and \( t_{p} \), \( v_{e} \) depends only
on \( \epsilon  \). For a particular choice of \( D=0.343 \) and \( t_{p}=8\times 10^{-4} \), Table 3 indicates the way
calculated values of \( v_{e} \) vary with \( \epsilon  \); the corresponding variation of
the ratio of \( v_{e} \) with \( v_{g} \) is also shown in Table 3. Of course, more detailed
studies are called for in order to have a precise quantitative idea
about the dependence of \( v_{e} \) on the relevant parameters. 

An important point to note from Table 3 is that \( v_{e} \) can exceed \( v_{g} \) \em substantially\em .
It should therefore be interesting to investigate whether any \em bound
\em exists on the ratio of \( v_{e} \) with \( v_{g} \). It may also be worthwhile to
compute our example for various cases by varying mass and width of
the wave packet to see what happens to this ratio in the various limiting
situations such as for large mass (classical limit) and for broad
wave packets (plane wave limit). \\
To sum up, our work serves to reveal that much interesting and counterintuitive
physics is concealed \em within \em the time evolution of reflection/transmission
probability for a wave packet scattered from a perturbed barrier.
This has so far remained unexplored because attention is usually focused
\em only \em on the final time independent values of reflection/transmission
probability. In particular, the effects uncovered in this paper involve
an intriguing interplay between ``particle'' and ``wave'' aspects of a
wave packet, its conceptual ramifications warranting further probing.
Such effects could also be amenable to experimental verification using
the available neutron/electron ``single particle'' experimental arrangements
[7,8].

We are grateful to Shyamal Sengupta for his stimulating comments motivating
this work. Thanks are due to Girish agarwal for his useful suggestions.
We also thank Sayandeb Basu, Guruprasad Kar and Manoj Samal for helpful
discussions. One of the authors S.B. likes to thank Sibaji Banerjee
for his useful suggestions on computation. D. H. acknowledges the
support provided by the Dept. of Science and Technology, Govt. of
India.

\( \smallskip  \)

[1] B. M. Garraway and K-A. Souminen, Rep. Prog. Phys. \bfseries 58\mdseries ,
365 (1995); W. S. Warren, H. Rabitz, and M. Dahleh, Science \bfseries 259\mdseries ,
1581 (1993); M. J. J. Vrakking, D. M. Villeneueve, and Albert Stolow,
Phys. Rev. A \bfseries 54\mdseries , 37 (1996). 

[2] I. Sh. Averbukh, M. J. J. Vrakking, D.M. Villeneueve, and Albert
Stolow, Phys. Rev. Lett. \bfseries 77\mdseries , 3518 (1996).

[3] R. Bluhm, V.A. Kostelecky, and J.A. Porter, Am. J. Phys. \bfseries 64\mdseries ,
944 (1996). 

[4] M.V. Berry, J. Phys. A \bfseries 29\mdseries , 6617 (1996);
M.V. Berry and S. Klein, J. Mod. Opt. \bfseries 43\mdseries ,
2139 (1996). 

[5] D.L. Aronstein and C.R. Stroud, Phys. Rev. A \bfseries 55\mdseries ,
4526 (1997). 

[6] A. Venugopalan and G.S. Agarwal, Phys. Rev. A \bfseries 59\mdseries ,
1413 (1999). 

[7] A. Tonomura, J. Endo, T. Matsuda, T. Kawasaki and H. Ezawa, Am.
J. Phys., \bfseries 57 \mdseries (2), 117 (1989); S. A. Werner
in \em Fundamental Problems in Quantum Theory, \em eds. D. M. Greenberger
and A. Zeilinger (N. Y. Academy of Science, 1995), p. 241. 

[8] H. Rauch, in \em Fundamental Problems in Quantum Theory, \em eds.
D. M. Greenberger and A. Zeilinger (N. Y. Academy of Science, 1995),
p. 263. 

[9] A. Goldberg, H. M. Schey and J. L. Schwartz, Am. J. Phys. \bfseries 35
\mdseries , 177 (1967).

[10] D. M. Greenberger, \em \em Review of Modern Physics, \bfseries 55\mdseries ,
875 (1983).

[11] J. S. Bell, \em \em Physics, \bfseries 1\mdseries , 195
(1964). For a recent overview see, for example, D. Home, \em Conceptual
Foundations of Quantum Physics - An Overview from Modern Perspectives
\em (Plenum Press, New York, 1997), Chapter 4 and references there
in.

\( \pagebreak  \)

\bfseries \hfill{}Table Captions\hfill{} \mdseries 

\bfseries Table 1: \mdseries Magnitudes of the time interval
(\( \Delta t \)) over which ``superarrivals `` occur and the measure \( (\eta ) \) of ``superarrivals''.
Note that \( \Delta t \) is \em almost independent \em of the span of perturbation
\( \epsilon  \) whereas \( \eta  \) has an explicit \em dependence \em on \( \epsilon  \).

\bfseries Table 2: \mdseries \em Violation \em of the locality
condition \( t_{d}>\tau  \).

\bfseries Table 3: \mdseries Velocity \( v_{e} \) with which the effect
of perturbation propagates across the wave packet is substantially
\em higher \em than the group velocity  \( v_{g} \). Note that \( v_{e}/v_{g} \) is \em independent
\em of the span of perturbation.

\( \medskip  \)

\bfseries \hfill{}Figure Captions\hfill{} \mdseries 

\bfseries Figure 1: \mdseries Profile of the wave packet at
t=0.

\bfseries Figure 2: \mdseries The time evolution of reflection
probability in the unperturbed situation. Note that the curve gradually
tends towards its asymptotic (time-independent) value.

\bfseries Figure 3: \mdseries Profile of the wave packet at
\( t=t_{p} \). Overlap of the wave packet with the barrier is crucial for the
effect of ``superarrivals''.

\bfseries Figure 4: \mdseries The time evolution of \( \left| R_{p}(t)\right| ^{2} \) for various
magnitudes of \( N(or,\epsilon  \)). 

\bfseries Figure 5: \mdseries A comparison between \( \left| R_{s}(t)\right| ^{2} \) and \( \left| R_{p}(t)\right| ^{2} \) for
\( N=2 \).

\bfseries Figure 6: \mdseries Profiles of the wave packet at
different times of perturbation.

\( \pagebreak  \)

\hfill{} \bfseries TABLE 1\mdseries \hfill{} 

\vspace{0.30cm}
{\centering \begin{tabular}{|c|c|c|c|c|c|}
\hline 
\( N \)&\( \epsilon \times 10^{-3} \)&\( t_{d}\times 10^{-3} \)&\( t_{c}\times 10^{-3} \)&\( \Delta t=t_{c}-t_{d} \)&\( \eta  \)\\
\hline 
\hline 
\( 2 \)&\( 0.004 \)&\( 1.122 \)&\( 1.832 \)&\( 0.71 \)&\( 0.50 \)\\
\hline 
\( 10 \)&\( 0.02 \)&\( 1.114 \)&\( 1.828 \)&\( 0.714 \)&\( 0.46 \)\\
\hline 
\( 30 \)&\( 0.06 \)&\( 1.094 \)&\( 1.814 \)&\( 0.72 \)&\( 0.37 \)\\
\hline 
\( 50 \)&\( 0.1 \)&\( 1.072 \)&\( 1.792 \)&\( 0.72 \)&\( 0.28 \)\\
\hline 
\end{tabular}\par}
\vspace{0.30cm}

\( \smallskip  \)

\hfill{} \bfseries TABLE 2\mdseries \hfill{} 

\vspace{0.30cm}
{\centering \begin{tabular}{|c|c|c|c|c|}
\hline 
\( N \)&\( \epsilon \times 10^{-3} \)&\( t_{d}\times 10^{-3} \)&\( \tau \times 10^{-3} \)&locality condition\\
\hline 
\hline 
2&\( 0.004 \)&\( 1.122 \)&\( 1.890 \)&violated\\
\hline 
10&\( 0.02 \)&\( 1.114 \)&\( 1.882 \)&do\\
\hline 
30&\( 0.06 \)&\( 1.094 \)&\( 1.862 \)&do\\
\hline 
50&\( 0.1 \)&\( 1.072 \)&\( 1.842 \)&do\\
\hline 
\end{tabular}\par}
\vspace{0.30cm}

\( \smallskip  \)

\hfill{} \bfseries TABLE 3 \mdseries \hfill{} 

\vspace{0.30cm}
{\centering \begin{tabular}{|c|c|c|c|c||c||c|}
\hline 
\( N \)&\( \epsilon \times 10^{-3} \)&\( t_{d}\times 10^{-3} \)&\( \tau \times 10^{-3} \)&\( v_{e} \)&\( v_{0}=2p_{0} \)&\( \frac{v_{e}}{v_{0}} \)\\
\hline 
\hline 
2&\( 0.004 \)&\( 1.122 \)&\( 1.890 \)&337.15\( \pi  \)&100\( \pi  \)&3.37\\
\hline 
10&\( 0.02 \)&\( 1.114 \)&\( 1.882 \)&337.15\( \pi  \)&do&3.37\\
\hline 
30&\( 0.06 \)&\( 1.094 \)&\( 1.862 \)&337.15\( \pi  \)&do&3.37\\
\hline 
50&\( 0.1 \)&\( 1.072 \)&\( 1.842 \)&339.24\( \pi  \)&do&3.39\\
\hline 
\end{tabular}\par}
\vspace{0.30cm}

\( \smallskip  \)

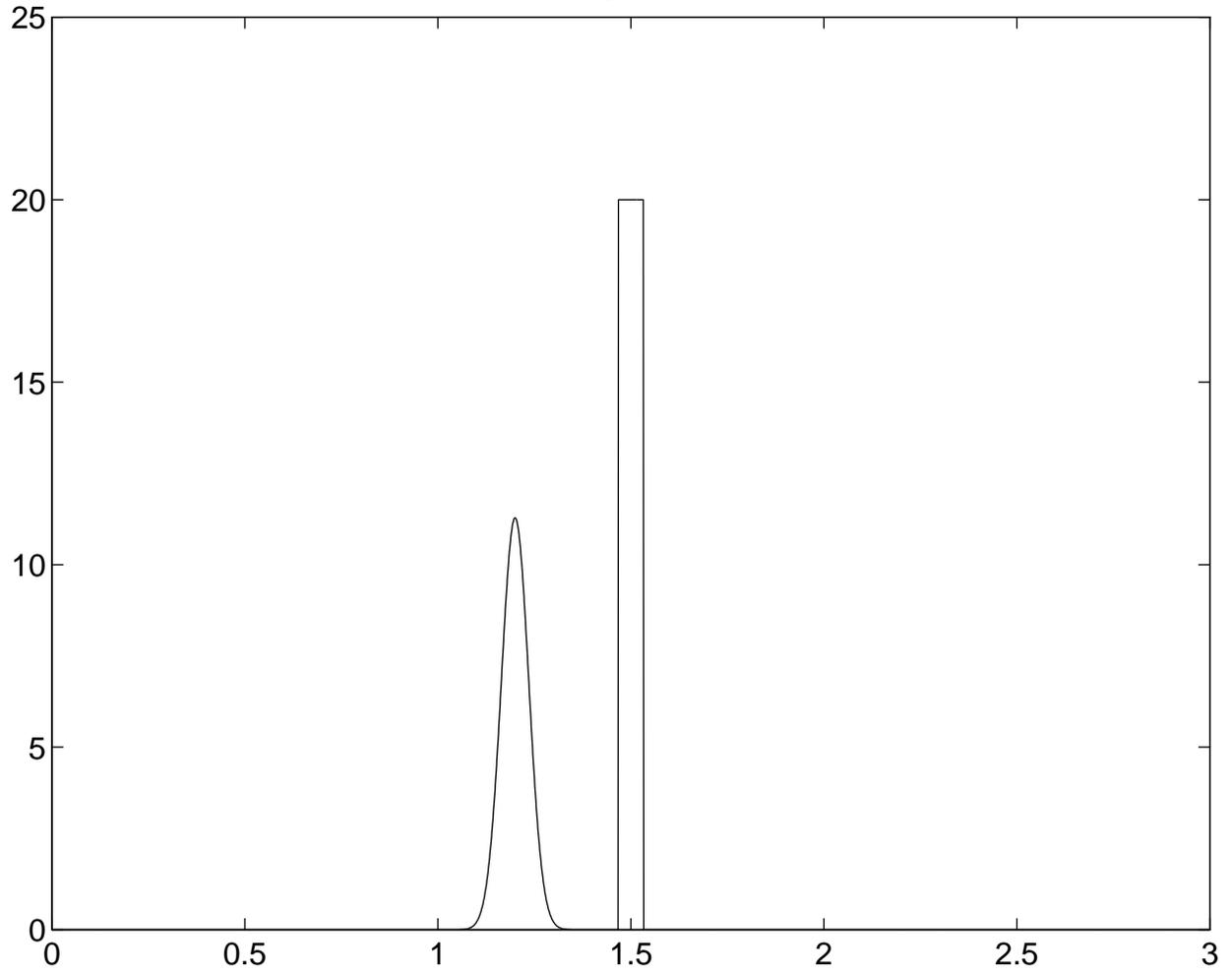
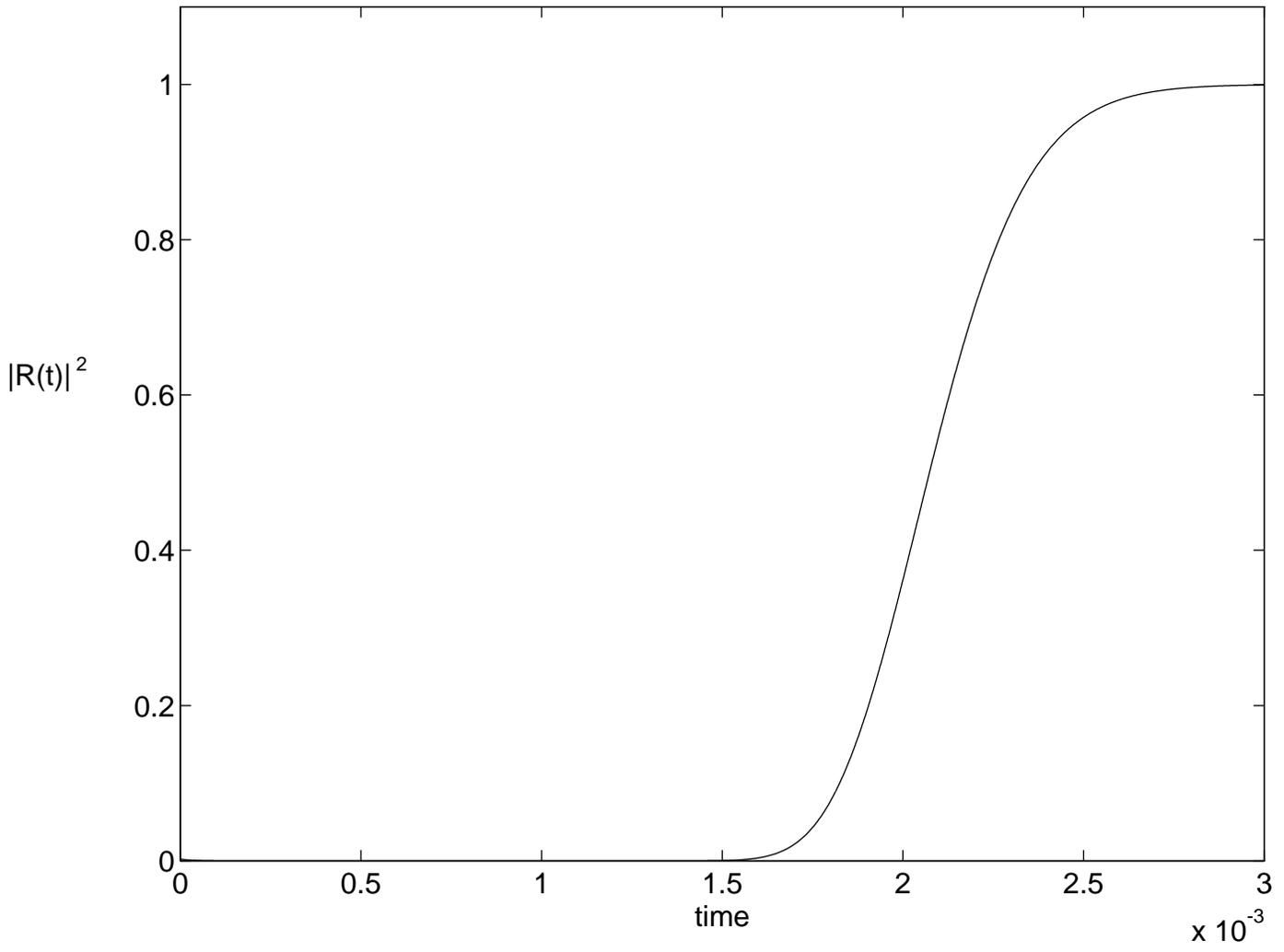
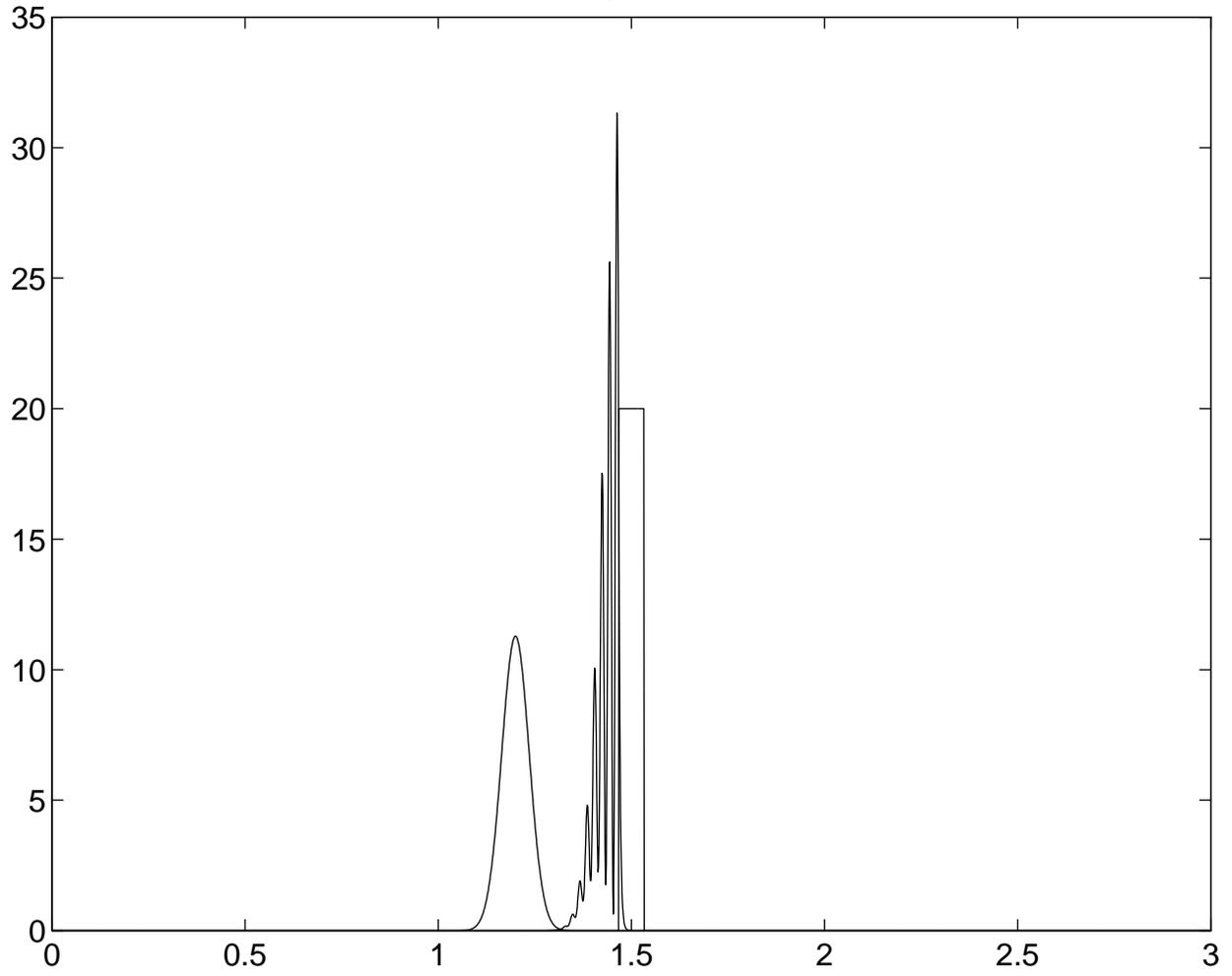
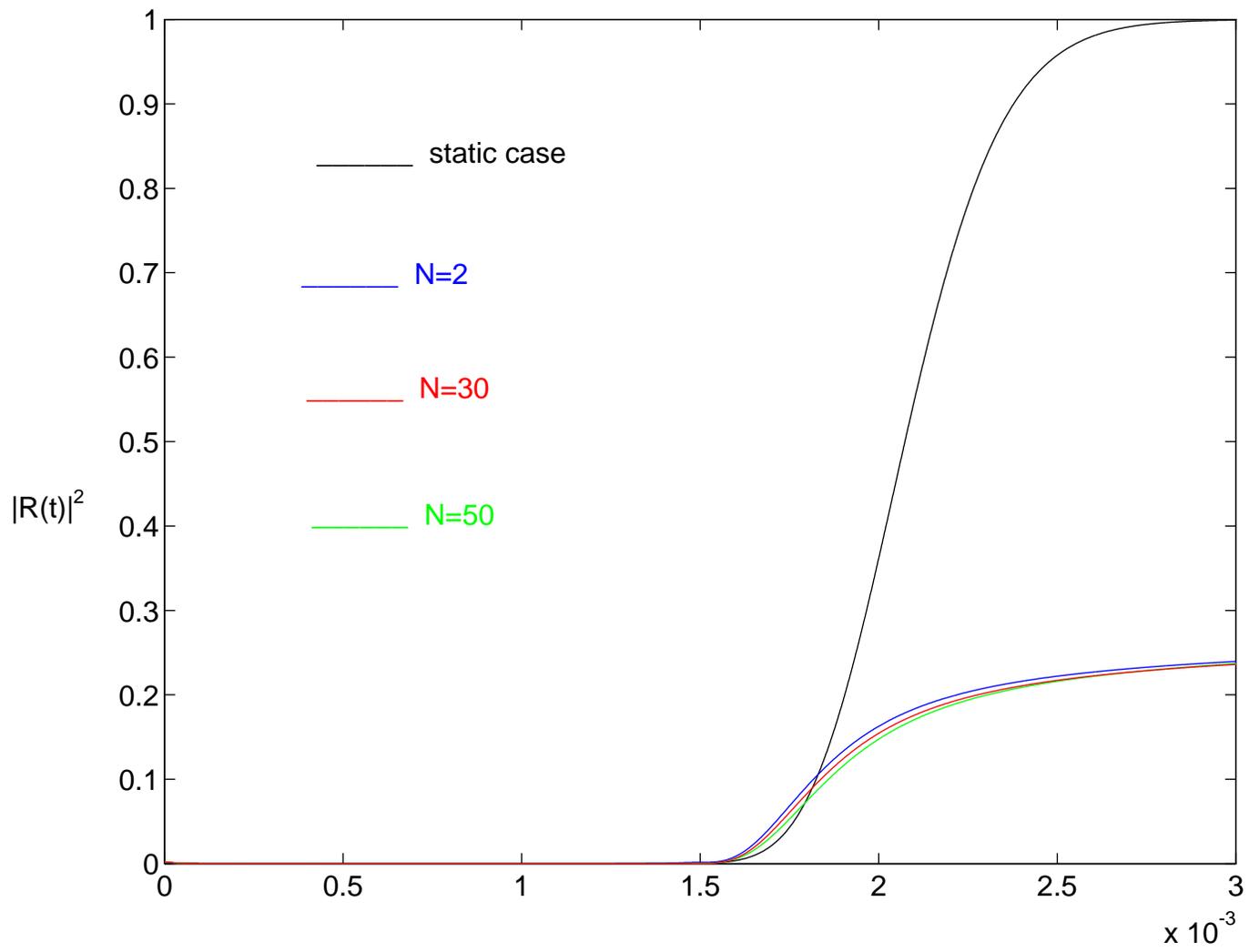
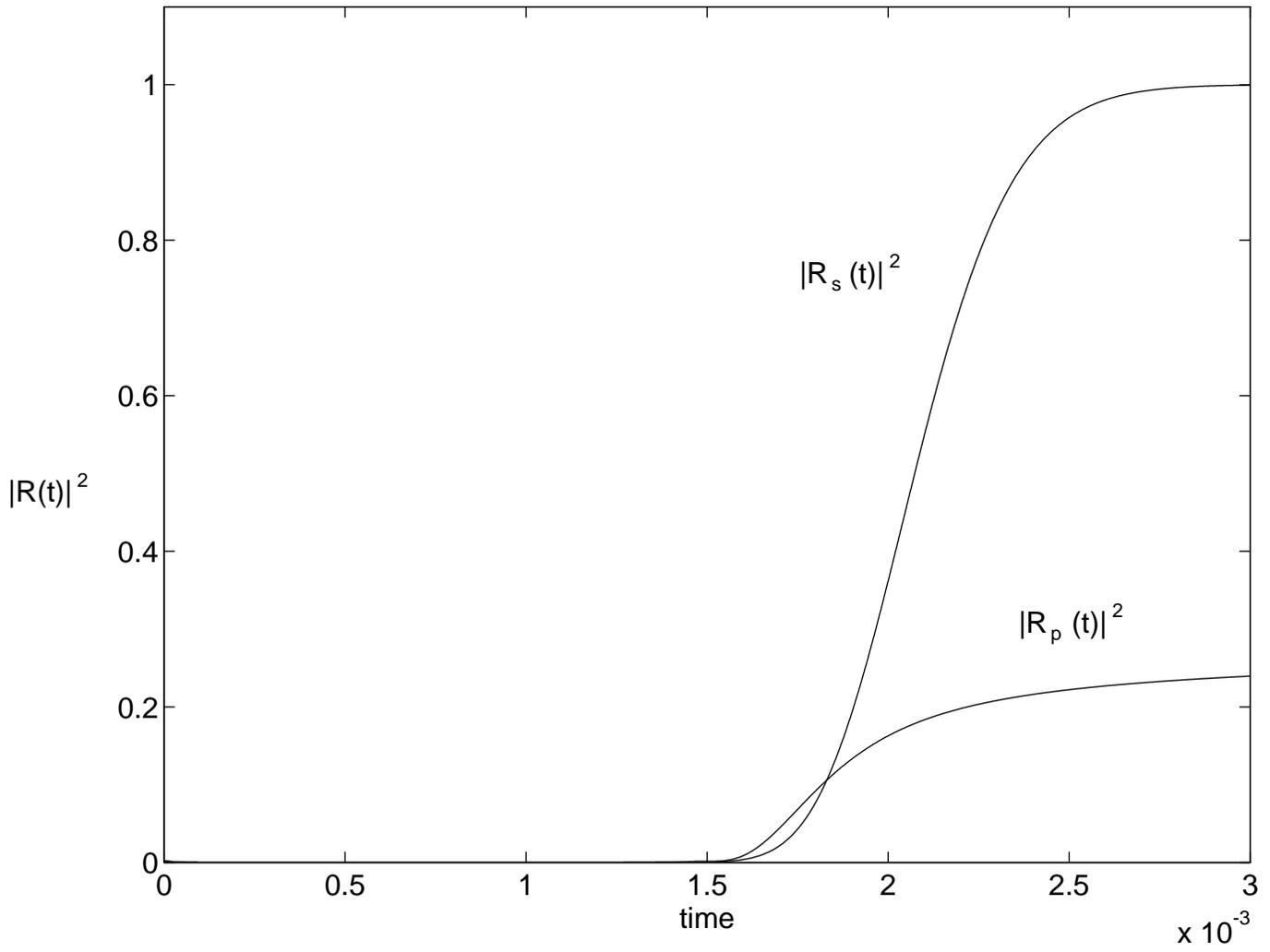
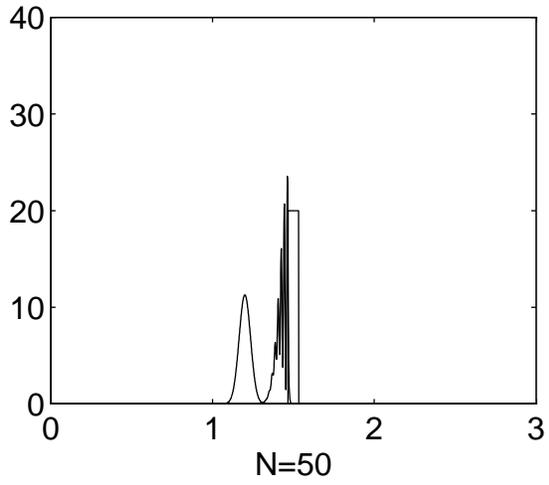
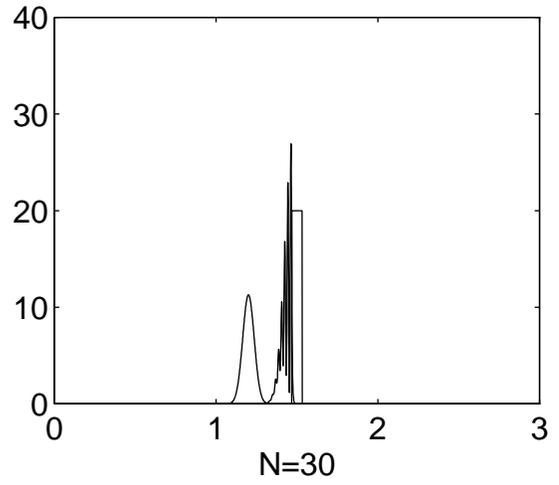
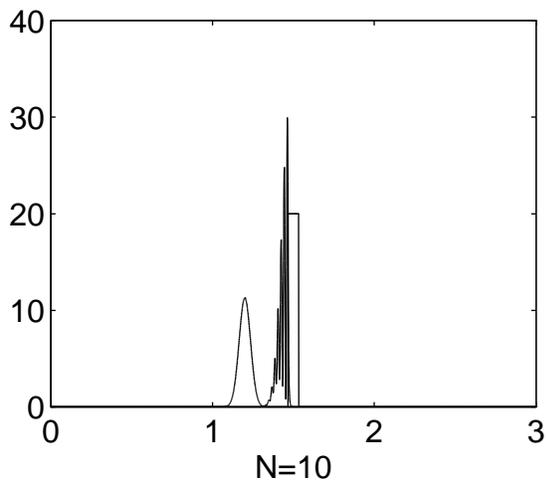
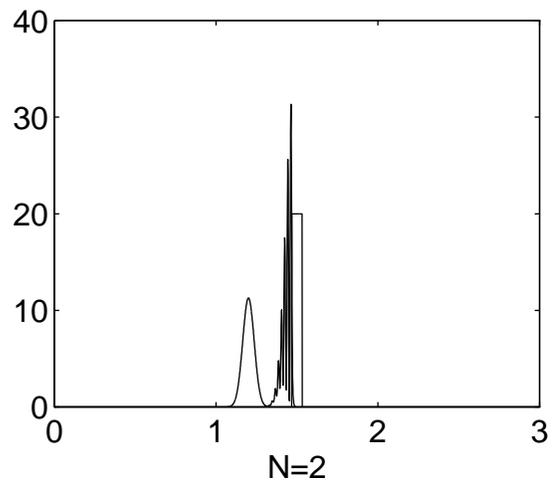
\end{document}